# Unified treatment of a class of spherically symmetric potentials: quasi-exact solution

*H. Panahi [1] and  M. Baradaran[2]*

*Department of Physics, University of Guilan, Rasht 41635-1914, Iran*

**Abstract**

In this paper, we investigate the Schrödinger equation for a class of spherically symmetric potentials in a simple and unified manner using the Lie algebraic approach within the framework of quasi-exact solvability. We illustrate that all models give rise to the same basic differential equation, which is expressible as an element of the universal enveloping algebra of $sl(2)$. Then, we obtain the general exact solutions of the problem by employing the representation theory of $sl(2)$ Lie algebra.

## 1. Introduction

From the viewpoint of solvability, the spectral problems are divided into two main classes, the exactly solvable (ES) models and exactly non-solvable models. A quantum model is called ES if for all its energy levels and corresponding wavefunctions, explicit expressions can be determined algebraically. These models are distinguished by the fact that there is a natural basis in the Hilbert space in which the infinite-dimensional Hamiltonian can be diagonalized with the help of algebraic methods. In the literature, considerable efforts have been devoted to obtaining the exact solutions of the relativistic and non-relativistic equations using different methods and techniques [1-5]. In contrast, the exactly non-solvable models are the spectral problems which their infinite-dimensional Hamiltonian cannot be diagonalized algebraically. Unfortunately, there are only a small number of potentials for which the Schrödinger equation can be solved exactly, such as the harmonic oscillator [6,7], Pöschl-Teller [8,9], Coulomb [10-12], Morse [13,14], Rosen-Morse [15], Manning-Rosen [16,17], Tietz [18], etc [19-22]. In 1980's, the series of papers by Shifman, Ushveridze and Turbiner was devoted to the introduction of an intermediate class between the ES and the exactly non-solvable models for which a certain finite number of eigenvalues and eigenfunctions, but not the whole spectrum, can be calculated exactly by algebraic methods. They were called quasi-exactly solvable (QES) [23-26]. These models are distinguished by the fact that the Hamiltonian is expressible as a quadratic combination of the generators of a finite-dimensional Lie algebra of first order differential operators preserving a finite-dimensional subspace of functions and thereby can be represented as a block-diagonal matrix with at least one finite block. Thus, the problem reduces to diagonalizing this block and computing the corresponding eigenvalues and eigenfunctions, which always can be done. In this paper, using the Lie algebraic approach, we present a simple unified derivation and exact solution of the Schrödinger equation for a class of four spherically symmetric potentials within the framework of quasi-exact solvability. We demonstrate that all four cases are reducible to the same basic differential equation which can be solved exactly due to the existence of a hidden $sl(2)$ symmetry. Then, with the aid of the representation theory of $sl(2)$, the general exact solution to the basic equation is determined.

---

[1] Corresponding author E-mail:t-panahi@guilan.ac.ir

[2] E-mail: marzie.baradaran@yahoo.com

This paper is organized as follows: in section 2, we briefly review the Lie algebraic approach of quasi-exact solvability. In section 3, we introduce the four systems and transform the corresponding equations into a same form that is suitable for $sl(2)$ algebraization and can be expressed as an element of the universal enveloping algebra of $sl(2)$. Then, using the representation theory of $sl(2)$, we obtain the general exact solution to the basic equation in section 4. Also, the closed-form expressions for the eigenvalues and eigenfunctions as well as the allowed parameters of potentials are given for each of the systems. We end with conclusions in section 5.

## 2. Quasi-exact solvability through the *sl*(2) algebraization

The general problem of quantum mechanics is to solve the Schrödinger equation $H\Psi = E\Psi$, where the wavefunction $\Psi$ belongs to the space of square integrable functions $L^2(R)$. An eigenvalue equation is called QES if there is a non-trivial finite-dimensional subspace $U$ of $L^2(R)$ which is invariant under $H$, i.e. $HU \subseteq U$ [25,26]. According to Ref. [25], any one-dimensional QES differential equation possesses a hidden Lie algebra $sl(2)$ which is the only Lie algebra of first-order differential operators that possesses a finite-dimensional representation. This implies that they can be rewritten in terms of $sl(2)$ generators with differential operators [23-25]

$$J_n^+ = -x^2 \frac{d}{dx} + nx,$$

$$J_n^0 = x\frac{d}{dx} - \frac{n}{2}, \qquad (1)$$

$$J_n^- = \frac{d}{dx},$$

which obey the commutation relations

$$[J_n^+, J_n^-] = 2J_n^0, \quad [J_n^0, J_n^\pm] = \pm J_n^\pm, \qquad (2)$$

and leave invariant the $(n+1)$-dimensional vector space of polynomials

$$P_{n+1}(x) = \langle 1, x, x^2, \ldots, x^n \rangle. \qquad (3)$$

With these properties, it is easy to verify that the most general second order differential operator in the enveloping algebra of $sl(2)$ takes the form

$$H = \sum_{a,b=0,\pm} C_{ab} J^a J^b + \sum_{a=0,\pm} C_a J^a + C, \qquad (4)$$

where $C_{ab}$, $C_a$ and $C$ are real constants. On the other hand, this operator as an ordinary differential equation has the general form

$$H\phi(x) = 0,$$

$$H = -P_4(x)\frac{d^2}{dx^2} + P_3(x)\frac{d}{dx} + P_2(x), \qquad (5)$$

where $P_j$ are the polynomials of degree $j$. This operator can be turned into the Schrödinger-like operator

$$\tilde{H} = e^{-A(z)} H e^{A(z)} = -\frac{1}{2}\frac{d^2}{dz^2} + (A')^2 - A'' + P_2(x(z)), \qquad (6)$$

through the following change of variable and gauge transformation

$$z = \pm \int \frac{dx}{\sqrt{P_4}},$$

$$\phi(x) = e^{-\int \left(\frac{P_3}{P_4}\right)dx + \log z'} \psi(z). \qquad (7)$$

In the next section, using the method given above, we show that exact solutions of the Schrödinger equation for the four models can be simply obtained in a unified treatment.

## 3. The four models and the corresponding differential equations

In this section, we introduce the models which will be the object of study in this paper. For each case, we illustrate that the corresponding Schrödinger equation is reducible to the basic differential equation of second order which is QES due to the existence of a hidden $sl(2)$ algebraic structure.

### 3.1. Non-polynomial Potential

First, we consider the non-polynomial oscillator defined as [27]

$$V(r) = r^2 + \frac{\alpha r^2}{1+\beta r^2}, \tag{8}$$

where $\beta > 0$ and $\alpha$ is a real constant. This potential appears in various branches of physics such as the zero-dimensional quantum field theory with nonlinear Lagrangian [28,29], quantum mechanics [30,31], laser physics [32,33], etc. This potential has been studied by a variety of methods including the analytic continued fractions [34], the supersymmetric quantum mechanics [35], the $1/N$ expansion method [27], the wavefunction ansatz method [12] and the Bethe ansatz method [36]. In atomic units ($m = \hbar = c = 1$), the radial Schrödinger equation with potential (8) is

$$\left(-\frac{1}{2}\frac{d^2}{dr^2} + \frac{l(l+1)}{2r^2} + r^2 + \frac{\alpha r^2}{1+\beta r^2} - E\right)\psi(r) = 0. \tag{9}$$

Using the change of variable $z = -\beta r^2$ and making the gauge transformation

$$\psi(z) = (1-z)z^{\frac{l+1}{2}} e^{-\frac{z}{\beta\sqrt{2}}} \varphi(z), \tag{10}$$

which preserves the asymptotic behaviour of the wavefunction at the origin and infinity, Eq. (9) becomes

$$\left\{z(z-1)\frac{d^2}{dz^2} + \left(-\frac{2}{\beta\sqrt{2}}z^2 + (\frac{2}{\beta\sqrt{2}} + l + \frac{7}{2})z - (l+\frac{3}{2})\right)\frac{d}{dz}\right.$$
$$\left. + \left(-\frac{E}{2\beta} + \frac{\alpha}{2\beta^2} - \frac{1}{\beta\sqrt{2}}(l+\frac{7}{2})\right)z + \left(\frac{E}{2\beta} + \frac{l}{2} + 1 + \frac{1}{\beta\sqrt{2}}(l+\frac{3}{2})\right)\right\}\varphi(z) = 0. \tag{11}$$

### 3.2. Screened Coulomb Potential

Here, we consider the screened Coulomb potential defined by [37]

$$V(r) = \frac{\gamma}{r} + \frac{\delta}{r+\kappa}, \quad \gamma < -\delta. \tag{12}$$

The corresponding radial Schrödinger equation is given by

$$\left(-\frac{1}{2}\frac{d^2}{dr^2} + \frac{l(l+1)}{2r^2} + \frac{\gamma}{r} + \frac{\delta}{r+\kappa}\right)\psi(r) = E\psi(r). \tag{13}$$

Several methods and techniques for solving this problem can be found in Refs. [12, 36-39]. Applying the transformation

$$\psi(r) = (r+\kappa)r^{l+1}e^{-\sqrt{-2E}(r+\kappa)}\varphi(r), \tag{14}$$

and also replacing the variable $r$ by $z$, we obtain

$$\left\{z(z+\kappa)\frac{d^2}{dz^2} + 2\left(-\sqrt{-2E}z^2 + (-\sqrt{-2E}\kappa + l + 2)z + (l+2)\kappa\right)\frac{d}{dz}\right.$$
$$\left. + \left(2(-\delta - \gamma - \sqrt{-2E}(l+2))z - 2(\kappa(l+1)\sqrt{-2E} + \kappa\gamma - l - 1)\right)\right\}\varphi(z) = 0. \tag{15}$$

### 3.3. Singular integer power potential

The problem of the singular power potentials, have been widely carried out in various branches of physics, in both classical and quantum mechanics [40-43]. Here, we consider the singular integer power potential as [44,45]

$$V(r) = \frac{\lambda}{r} + \frac{\mu}{r^2} + \frac{\xi}{r^3} + \frac{\tau}{r^4}, \tag{16}$$

with the corresponding Schrödinger equation given by

$$\left(-\frac{1}{2}\frac{d^2}{dr^2} + \frac{l(l+1)}{2r^2} + \frac{\lambda}{r} + \frac{\mu}{r^2} + \frac{\xi}{r^3} + \frac{\tau}{r^4}\right)\psi(r) = E\psi(r). \tag{17}$$

From the asymptotic behaviour of the wavefunction, we consider the following transformation

$$\psi(r) = r^{1+\frac{\xi}{\sqrt{2\tau}}} e^{-\left(\sqrt{-2E}r + \frac{\sqrt{2\tau}}{r}\right)} \varphi(r). \tag{18}$$

Substituting this into Eq. (17) and also replacing $r$ by $z$, we get

$$\left\{z^2\frac{d^2}{dz^2} + 2\left(-\sqrt{-2E}z^2 + (1+\frac{\xi}{\sqrt{2\tau}})z + \sqrt{2\tau}\right)\frac{d}{dz}\right.$$
$$\left. -2\left(\sqrt{-2E}(1+\frac{\xi}{\sqrt{2\tau}}) + \lambda\right)z - 2\mu - l(l+1) - 2\sqrt{-4\tau E} + \frac{\xi}{\sqrt{2\tau}} + \frac{\xi^2}{2\tau}\right\}\varphi(z) = 0. \tag{19}$$

### 3.4. Singular anharmonic potential

Here, following Ref. [46], we consider the potential

$$V(r) = \omega r^2 + \frac{\varepsilon}{r^2} + \frac{\sigma}{r^4} + \frac{\chi}{r^6}, \tag{20}$$

with the radial Schrödinger equation

$$\left(-\frac{1}{2}\frac{d^2}{dr^2} + \frac{l(l+1)}{2r^2} + \omega r^2 + \frac{\varepsilon}{r^2} + \frac{\sigma}{r^4} + \frac{\chi}{r^6}\right)\psi(r) = E\psi(r). \tag{21}$$

Similar to the previous cases, we extract the asymptotic behaviour of the wavefunction by making the following transformations

$$z = r^2,$$

$$\psi(r) = r^{\frac{3}{2} + \frac{\sigma}{\sqrt{2\chi}}} e^{-\left(\sqrt{\frac{\omega}{2}}r^2 + \sqrt{\frac{\chi}{2}}\frac{1}{r^2}\right)} \varphi(r). \tag{22}$$

After substituting this into Eq. (21), we obtain

$$\left\{z^2 \frac{d^2}{dz^2} + \left(-\sqrt{2\omega}z^2 + (2+\frac{\sigma}{\sqrt{2\chi}})z + \sqrt{2\chi}\right)\frac{d}{dz}\right.$$

$$\left. + \left(\frac{E}{2} - \sqrt{\frac{\omega}{2}}(2+\frac{\sigma}{\sqrt{2\chi}})\right)z - \frac{1}{4}\left(l'(l'+1) + 2\sqrt{4\omega\chi} - \frac{\sigma^2}{2\chi} - \frac{2\sigma}{\sqrt{2\chi}} - \frac{3}{4}\right)\right\}\varphi(z) = 0, \tag{23}$$

where

$$l' = \frac{-1 + \sqrt{4l^2 + 4l + 8\varepsilon + 1}}{2}. \tag{24}$$

### 4. Solutions of the basic differential equation for the four models

In the previous section, we have shown that our QES models, after the appropriate transformations, are expressible as second-order differential equations (11), (15), (19) and (23), respectively. These equations have the same basic structure

$$H\phi(z) = 0,$$

$$H = z(z-a)\frac{d^2}{dz^2} + \left(b_2 z^2 + b_1 z + b_0\right)\frac{d}{dz} + \left(c_1 z + c_0\right). \tag{25}$$

where $a$, $c_0$, $c_1$, $b_0$, $b_1$ and $b_2$ are real constants. Here, we intend to solve this equation using the Lie algebraic approach within the representation theory of $sl(2)$. More precisely,

from Eq. (4), the general form of a one-dimensional QES differential equation is as follows [25]

$$H = C_{++}J_n^+ J_n^+ + C_{+0}J_n^+ J_n^0 + C_{+-}J_n^+ J_n^- + C_{0-}J_n^0 J_n^- + C_{--}J_n^- J_n^- + C_{+}J_n^+ + C_{0}J_n^0 + C_{-}J_n^- + C, \quad (26)$$

which clearly preserves the ($n+1$)-dimensional representation space of the $sl(2)$ algebra as

$$\phi_n(z) = \sum_{m=0}^{n} p_m z^m, \quad n = 0, 1, 2, \ldots. \quad (27)$$

Comparing Eq. (25) with Eq. (26), it is seen that the differential operator $H$ can be expressed as a special case of the general form (26) as

$$H = -J_n^+ J_n^- - aJ_n^0 J_n^- - b_2 J_n^+ + (n+b_1)J_n^0 + \left(b_0 - \frac{n}{2}a\right)J_n^- + \left(\frac{n^2}{2} + \frac{n}{2}b_1 + c_0\right), \quad (28)$$

if the following constraint on the coefficients holds

$$c_1 = -nb_2. \quad (29)$$

Hence, we have shown that the differential operator $H$ is an element of the universal enveloping algebra of $sl(2)$ and thereby we can use the representation theory of $sl(2)$ to determine the solutions of the problem. In the ($n+1$)-dimensional space $\phi_n(z)$, the operators $J_n^+$, $J_n^0$ and $J_n^-$ can be represented by the $(n+1)\times(n+1)$ matrices in the basis $\langle 1, z, z^2, \ldots, z^{n+1} \rangle$

$$J_n^+ = \begin{pmatrix} 0 & 1 & 0 & \ldots & 0 & 0 \\ 0 & 0 & 2 & \ldots & 0 & 0 \\ \vdots & \vdots & \ldots & 3 & \vdots & \vdots \\ \vdots & \vdots & \ldots & \ldots & n-1 & 0 \\ 0 & 0 & \ldots & \ldots & 0 & n \\ 0 & 0 & \ldots & \ldots & 0 & 0 \end{pmatrix}, \quad J_n^- = \begin{pmatrix} 0 & 0 & 0 & \ldots & 0 & 0 \\ n & 0 & 0 & \ldots & 0 & 0 \\ \vdots & n-1 & \ldots & \ldots & \vdots & \vdots \\ \vdots & \vdots & \ldots & \ldots & 0 & 0 \\ 0 & 0 & \ldots & 2 & 0 & 0 \\ 0 & 0 & \ldots & \ldots & 1 & 0 \end{pmatrix},$$

$$J_n^0 = \begin{pmatrix} \frac{n}{2} & 0 & \ldots & \ldots & 0 & 0 \\ 0 & \frac{n}{2}-1 & \ldots & \ldots & 0 & 0 \\ \vdots & \vdots & \ldots & \ldots & \vdots & \vdots \\ 0 & 0 & \ldots & \ldots & 1-\frac{n}{2} & 0 \\ 0 & 0 & \ldots & \ldots & 0 & -\frac{n}{2} \end{pmatrix}. \quad (30)$$

Substituting Eq. (30) into Eq. (28), results in a matrix equation whose non-trivial solution exists if the following condition is fulfilled (Cramer's rule)

$$\begin{vmatrix} c_0 & b_0 & 0 & 0 & 0 \\ -nb_2 & c_0+b_1 & 2b_0-2a & 0 & 0 \\ 0 & -(n-1)b_2 & c_0+2b_1+2 & \ddots & 0 \\ 0 & \ddots & \ddots & \ddots & \vdots \\ 0 & 0 & -2b_2 & c_0+(n-1)(b_1+n-2) & n(b_0-(n-1)a) \\ 0 & 0 & 0 & -b_2 & c_0+n(b_1+n-1) \end{vmatrix} = 0. \qquad (31)$$

This equation yields a $(n+1)$-set of important constraints on the potential parameters that vary for different $n$. These constraints together with Eq. (29), yields the exact solutions of the systems. It can be easily shown that the expansion coefficients $p_m$'s in Eq. (27) satisfy the three-term recursion relation

$$p_{m+1} = \frac{(2b_2)p_{m-1} - (c_0 + m(b_1 + m - 1))p_m}{(m+1)(b_0 - ma)}, \qquad (32)$$

with the boundary conditions $p_{n+1} = 0$ and $p_{-1} = 0$. Therefore, we have succeeded in obtaining the exact expressions for the energies, wavefunctions and the allowed values of the potential parameters for the $n+1$ first states algebraically. The main advantage of our algebraic method is that we can quickly obtain the general solutions of the systems for any arbitrary $n$ from Eqs. (27), (29) and (31) without the cumbersome numerical and analytical procedures usually involved in obtaining the solutions for higher states. In the following, we apply the above results to obtain explicit solutions for each of the four systems.

**4.1. Non-polynomial Potential**

In this case, from (11) with (25), we get

$$a = 1,$$
$$b_2 = -\frac{2}{\beta\sqrt{2}}, \qquad b_1 = \frac{2}{\beta\sqrt{2}} + l + \frac{7}{2}, \qquad b_0 = -(l + \frac{3}{2}), \qquad (33)$$
$$c_1 = -\frac{E}{2\beta} + \frac{\alpha}{2\beta^2} - \frac{1}{\beta\sqrt{2}}(l + \frac{7}{2}), \qquad c_0 = \frac{E}{2\beta} + \frac{l}{2} + 1 + \frac{1}{\beta\sqrt{2}}(l + \frac{3}{2}).$$

Then by Eq. (29), the closed form of the energy of the system is obtained as

$$E_n = \frac{\alpha}{\beta} - \sqrt{2}(2n + l + \frac{7}{2}), \qquad (34)$$

which, together with Eq. (31) yields the exact solutions of the system. Also, from Eqs. (10) and (27), the wavefunction of the model is obtained as

$$\psi_n(z) = (1-z) z^{\frac{l+1}{2}} e^{-\frac{z}{\beta\sqrt{2}}} \sum_{m=0}^{n} p_m z^m, \quad n = 0,1,2,..., \quad (35)$$

where the expansion coefficients $p_m$'s satisfy the recursion relation

$$p_{m+1} = \frac{\left(-\frac{4}{\beta\sqrt{2}}\right) p_{m-1} - \left(\frac{E}{2\beta} + \frac{l}{2} + 1 + \frac{1}{\beta\sqrt{2}}(l+\frac{3}{2}) + m\left(\frac{2}{\beta\sqrt{2}} + l + \frac{7}{2} + m - 1\right)\right) p_m}{-(m+1)\left((l+\frac{3}{2}) + m\right)}. \quad (36)$$

The results obtained for the first three states of this model are displayed in table 1.

**Table 1.** Exact solutions of the Schrödinger equation with non-polynomial Potential for the ground, first, and second excited states

| $n$ | Energy | Relation between potential parameters and $l$ | The radial wavefunction |
|---|---|---|---|
| 0 | $E_0 = \frac{\alpha}{\beta} - \sqrt{2}(l+\frac{7}{2})$ | $\frac{\alpha}{2\beta^2} - \frac{\sqrt{2}}{\beta} + \frac{l}{2} + 1 = 0$ | $\psi_0(r) = (1+\beta r^2)(-r\sqrt{\beta})^{l+1} e^{-\frac{r^2}{\sqrt{2}}} p_0$ |
| 1 | $E_1 = \frac{\alpha}{\beta} - \sqrt{2}(l+\frac{11}{2})$ | $((8l^2 + 46l + 53)\beta + 4E_1 l + 10E_1)\sqrt{2} + (6l^2 + 30l + 36)\beta^2$ $+ (8l+22)\beta(\frac{\alpha}{\beta} - \sqrt{2}(l+\frac{11}{2})) + 2E_1^2 + 4l^2 + 20l + 21 = 0$ | $\psi_1(r) = (1+\beta r^2)(-r\sqrt{\beta})^{l+1} e^{-\frac{r^2}{\sqrt{2}}}(p_0 - p_1\beta r^2),$ $p_1 = \frac{2}{2l+3}\left(\frac{E_1}{2\beta} + \frac{l}{2} + 1 + \frac{1}{\beta\sqrt{2}}(l+\frac{3}{2})\right) p_0$ |
| 2 | $E_2 = \frac{\alpha}{\beta} - \sqrt{2}(l+\frac{19}{2})$ | $\frac{1}{64}(2\sqrt{2}l + 2l\beta + 2E_2 + 3\sqrt{2} + 4\beta)^3$ $+ \frac{1}{32}(6l\beta + 25\beta + 6\sqrt{2})(2\sqrt{2}l + 2l\beta + 2E_2 + 3\sqrt{2} + 4\beta)^2$ $+ \left(2 + 4\sqrt{2}\beta(l+3) + \beta^2(l+\frac{9}{2})(l+\frac{7}{2})\right)\left((\sqrt{2}l + \frac{3}{\sqrt{2}}) + E_2 + (2+l)\beta\right)$ $+ (4l+6)\sqrt{2}\beta^2\left(\frac{9}{2} + \frac{\sqrt{2}}{\beta} + l\right)^{-1} = 0$ | $\psi_2(r) = (1+\beta r^2)(-r\sqrt{\beta})^{l+1} e^{-\frac{r^2}{\sqrt{2}}}(p_0 - p_1\beta r^2 + p_2\beta^2 r^4),$ $p_2 = \frac{1}{2l+5}\left(\left(\frac{4}{\beta\sqrt{2}}\right)p_0 + \left(\frac{E_2}{2\beta} + \frac{3l}{2} + \frac{1}{\beta\sqrt{2}}(l+\frac{7}{2}) + \frac{9}{2}\right)p_1\right),$ $p_1 = \frac{2}{2l+3}\left(\frac{E_2}{2\beta} + \frac{l}{2} + 1 + \frac{1}{\beta\sqrt{2}}(l+\frac{3}{2})\right) p_0$ |

## 4.2. Screened Coulomb Potential

In this case, comparing (15) with (25), we obtain

$$a = -\kappa,$$
$$b_2 = -2\sqrt{-2E}, \quad b_1 = 2\left(-\sqrt{-2E}\,\kappa + l + 2\right), \quad b_0 = (2l+4)\kappa, \tag{37}$$
$$c_1 = 2(-\delta - \gamma - \sqrt{-2E}\,(l+2)), \quad c_0 = -2(\kappa(l+1)\sqrt{-2E} + \kappa\gamma - l - 1).$$

Then by Eqs. (29) and (14), we obtain the following relations for the energy eigenvalues and the corresponding wavefunctions

$$E_n = \frac{-1}{2}\left(\frac{\delta+\gamma}{n+l+2}\right)^2, \tag{38}$$

$$\psi_n(r) = (r+\kappa)r^{l+1}e^{-\sqrt{-2E}\,(r+\kappa)}\sum_{m=0}^{n} p_m r^m, \quad n = 0,1,2,\ldots, \tag{39}$$

which, together with the determinant relation (31) give the exact solutions of this system. Also, the expansion coefficients $p_m$'s obey the three-term recursion relation

$$p_{m+1} = \frac{\left(-4\sqrt{-2E}\right)p_{m-1} - \left(-2(\kappa(l+1)\sqrt{-2E} + \kappa\gamma - l - 1) + m\left(2(-\sqrt{-2E}\,\kappa + l + 2) + m - 1\right)\right)p_m}{(m+1)(2l+4+m)\kappa}. \tag{40}$$

The results for the ground, first, and second excited states of this model are reported in table 2.

**Table 2.** Exact solutions of the Schrödinger equation with screened Coulomb Potential for the ground, first, and second excited states

| $n$ | Energy | Relation between potential parameters and $l$ | The radial wavefunction |
|---|---|---|---|
| 0 | $E_0 = \frac{-1}{2}\left(\frac{\delta+\gamma}{l+2}\right)^2$ | $\kappa(l+1)\sqrt{-2E_0} + \kappa\gamma - l - 1 = 0$ | $\psi_0(r) = (r+\kappa)r^{l+1}e^{-\sqrt{-2E_0}(r+\kappa)}p_0$ |
| 1 | $E_1 = \frac{-1}{2}\left(\frac{\delta+\gamma}{l+3}\right)^2$ | $(l+2)\kappa\sqrt{-2E_1}$ $+\left(-\sqrt{-2E_1}\,\kappa+l+2\right)\left(\kappa(l+1)\sqrt{-2E_1}+\kappa\gamma-l-1\right)$ $-\left(-\kappa(l+1)\sqrt{-2E_1}-\kappa\gamma+l+1\right)^2 = 0$ | $\psi_1(r) = (r+\kappa)r^{l+1}e^{-\sqrt{-2E_1}(r+\kappa)}(p_0+p_1 r),$ $p_1 = \frac{\left(+2(\kappa(l+1)\sqrt{-2E_1}+\kappa\gamma-l-1)\right)}{(2l+4)\kappa}p_0$ |
| 2 | $E_2 = \frac{-1}{2}\left(\frac{\delta+\gamma}{l+4}\right)^2$ | $\left(-\kappa(l+1)\sqrt{-2E_2}-\kappa\gamma+l+1\right)^3$ $+3\left(\kappa(l+1)\sqrt{-2E_2}+\kappa\gamma-l-1\right)^2\left(-\sqrt{-2E_2}\,\kappa+l+\frac{7}{3}\right)$ $-2\left(\kappa(l+1)\sqrt{-2E_2}+\kappa\gamma-l-1\right)\left(-2E_2\kappa^2+(-4\kappa l-9\kappa)\sqrt{-2E_2}+\frac{9l}{2}+5+l^2\right)$ $-2\kappa(l+2)\sqrt{-2E_2}\left(5-2\sqrt{-2E_2}\,\kappa+2l\right) = 0$ | $\psi_2(r) = (r+\kappa)r^{l+1}e^{-\sqrt{-2E_2}(r+\kappa)}(p_0+p_1 r+p_2 r^2),$ $p_2 = \frac{\left(2\sqrt{-2E_2}\right)p_0 - \left((l+2)\kappa\sqrt{-2E_2}+\kappa\gamma-2l-3\right)p_1}{(-2l-5)\kappa},$ $p_1 = \frac{\left(2(\kappa(l+1)\sqrt{-2E_2}+\kappa\gamma-l-1)\right)}{(2l+4)\kappa}p_0$ |

### 4.3. Singular integer power potential

In this case, from (19) and (25), we have

$$a = 0,$$
$$b_2 = -2\sqrt{-2E}, \qquad b_1 = 2(1 + \frac{\xi}{\sqrt{2\tau}}), \qquad b_0 = 2\sqrt{2\tau}, \qquad (41)$$
$$c_1 = -2\left(\sqrt{-2E}\,(1 + \frac{\xi}{\sqrt{2\tau}}) + \lambda\right), \qquad c_0 = -2\mu - l(l+1) - 2\sqrt{-4\tau E} + \frac{\xi}{\sqrt{2\tau}} + \frac{\xi^2}{2\tau}.$$

Then from Eqs. (29) and (18), we obtain the following relations for energy and wavefunction

$$E_n = \frac{-\lambda^2}{2}\left(n + \frac{\xi}{\sqrt{2\tau}} + 1\right)^{-2}, \qquad (42)$$

$$\psi_n(r) = r^{1+\frac{\xi}{\sqrt{2\tau}}} e^{-\left(\sqrt{-2E}\,r + \frac{\sqrt{2\tau}}{r}\right)} \sum_{m=0}^{n} p_m r^m, \qquad n = 0, 1, 2, \ldots, \qquad (43)$$

where the expansion coefficients $p_m$'s satisfy the following recursion relation

$$p_{m+1} = \frac{\left(-4\sqrt{-2E}\right)p_{m-1} - \left(-2\mu - l(l+1) - 2\sqrt{-4\tau E} + \frac{\xi}{\sqrt{2\tau}} + \frac{\xi^2}{2\tau} + m\left(2(1 + \frac{\xi}{\sqrt{2\tau}}) + m - 1\right)\right)p_m}{2(m+1)\sqrt{2\tau}}. \qquad (44)$$

These relations, together with Eq. (31) yield the exact solutions of the system. The results determined for the ground, first, and second excited states are displayed in table 3.

**Table 3.** Exact solutions of the Schrödinger equation with singular integer power potential for the ground, first, and second excited states

| $n$ | Energy | Relation between potential parameters and $l$ | The radial wavefunction |
|---|---|---|---|
| 0 | $E_0 = \dfrac{-\lambda^2}{2\left(\dfrac{\xi}{\sqrt{2\tau}}+1\right)^2}$ | $-2\mu - l(l+1) - 2\sqrt{-4\tau E_0} + \dfrac{\xi}{\sqrt{2\tau}} + \dfrac{\xi^2}{2\tau} = 0$ | $\psi_0(r) = r^{1+\frac{\xi}{\sqrt{2\tau}}} e^{-\left(\sqrt{-2E_0}\,r + \frac{\sqrt{2\tau}}{r}\right)} p_0$ |
| 1 | $E_1 = \dfrac{-\lambda^2}{2\left(\dfrac{\xi}{\sqrt{2\tau}}+2\right)^2}$ | $-4\sqrt{2}\sqrt{\tau}\sqrt{-2E_1}$ $+\dfrac{1}{2\tau}\left(\xi\sqrt{2}\sqrt{\tau} - 2l(l+1)\tau - 8\sqrt{-\tau E_1}\tau - 4\tau\mu + \xi^2\right)\left(2+\dfrac{\xi\sqrt{2}}{\sqrt{\tau}}\right)$ $+\left(-2\mu - l(l+1) - 4\sqrt{-\tau E_1} + 1/2\dfrac{\xi\sqrt{2}}{\sqrt{\tau}} + 1/2\dfrac{\xi^2}{\tau}\right)^2 = 0$ | $\psi_1(r) = r^{1+\frac{\xi}{\sqrt{2\tau}}} e^{-\left(\sqrt{-2E_1}\,r + \frac{\sqrt{2\tau}}{r}\right)} (p_0 + p_1 r),$ $p_1 = -\dfrac{1}{2\sqrt{2\tau}}\left(-2\mu - l(l+1) - 2\sqrt{-4\tau E_1} + \dfrac{\xi}{\sqrt{2\tau}} + \dfrac{\xi^2}{2\tau}\right) p_0$ |
| 2 | $E_2 = \dfrac{-\lambda^2}{2\left(\dfrac{\xi}{\sqrt{2\tau}}+3\right)^2}$ | $\left(-2\mu - l(l+1) - 4\sqrt{-\tau E_2} + \dfrac{\xi}{2}\sqrt{\dfrac{2}{\tau}} + \dfrac{1}{2}\dfrac{\xi^2}{\tau}\right)^3$ $+\dfrac{3}{4\tau^2}\left(\xi\sqrt{\dfrac{2}{\tau}}+\dfrac{8}{3}\right)\left(\xi\sqrt{2\tau} - 2l(l+1)\tau - 8\sqrt{-\tau E_2}\tau - 4\tau\mu + \xi^2\right)^2$ $+\left(\dfrac{2\xi^2}{\tau^2} + \dfrac{5\xi}{\tau}\sqrt{\dfrac{2}{\tau}} - 16\sqrt{\dfrac{-E_2}{\tau}} + \dfrac{6}{\tau}\right)\left(\xi\sqrt{2\tau} - 2l(l+1)\tau - 8\tau\sqrt{-\tau E_2} - 4\tau\mu + \xi^2\right)$ $-16\sqrt{2\tau}\sqrt{-2E_2}\left(3+\dfrac{\xi\sqrt{2}}{\sqrt{\tau}}\right) = 0$ | $\psi_2(r) = r^{1+\frac{\xi}{\sqrt{2\tau}}} e^{-\left(\sqrt{-2E_2}\,r + \frac{\sqrt{2\tau}}{r}\right)} (p_0 + p_1 r + p_2 r^2),$ $p_2 = \dfrac{\left(-\sqrt{-2E_2}\right) p_0 + \dfrac{1}{4}\left(2\mu + l(l+1) + 2\sqrt{-4\tau E_2} - \dfrac{3\xi}{\sqrt{2\tau}} - \dfrac{\xi^2}{2\tau} - 2\right) p_1}{\sqrt{2\tau}}$ $p_1 = -\dfrac{1}{2\sqrt{2\tau}}\left(-2\mu - l(l+1) - 2\sqrt{-4\tau E_2} + \dfrac{\xi}{\sqrt{2\tau}} + \dfrac{\xi^2}{2\tau}\right) p_0$ |

### 4.4. Singular anharmonic potential

In this case, from (23) and (25), we get

$$a = 0,$$
$$b_2 = -\sqrt{2\omega}, \qquad b_1 = \left(2 + \dfrac{\sigma}{\sqrt{2\chi}}\right), \qquad b_0 = \sqrt{2\chi}, \tag{45}$$
$$c_1 = \dfrac{E}{2} - \sqrt{\dfrac{\omega}{2}}\left(2 + \dfrac{\sigma}{\sqrt{2\chi}}\right), \qquad c_0 = -\dfrac{1}{4}\left(l'(l'+1) + 2\sqrt{4\omega\chi} - \dfrac{\sigma^2}{2\chi} - \dfrac{2\sigma}{\sqrt{2\chi}} - \dfrac{3}{4}\right).$$

Then from Eqs. (29) and (22), we obtain

$$E_n = \sqrt{2\omega}\left(2n + \dfrac{\sigma}{\sqrt{2\chi}} + 2\right), \tag{46}$$

$$\psi(r) = r^{\frac{3}{2}+\frac{\sigma}{\sqrt{2\chi}}} e^{-\left(\sqrt{\frac{\omega}{2}}r^2 + \sqrt{\frac{\chi}{2}}\frac{1}{r^2}\right)} \sum_{m=0}^{n} p_m r^m, \qquad n = 0, 1, 2, \ldots, \tag{47}$$

for the energy and wave function, respectively, which, together with Eq. (31) yield the exact solutions of the system. Also, the coefficients $p_m$'s satisfy the recursion relation

$$p_{m+1} = \frac{\left(-2\sqrt{2\omega}\right)p_{m-1} + \left(\frac{1}{4}\left(l'(l'+1) + 2\sqrt{4\omega\chi} - \frac{\sigma^2}{2\chi} - \frac{2\sigma}{\sqrt{2\chi}} - \frac{3}{4}\right) - m\left(2 + \frac{\sigma}{\sqrt{2\chi}}\right) + m - 1\right)p_m}{(m+1)\sqrt{2\chi}}. \qquad (48)$$

Solutions of the ground, first, and second excited states of this system are reported in table 4.

**Table 4.** Exact solutions of the Schrödinger equation with singular anharmonic potential for the ground, first, and second excited states

| $n$ | Energy | Relation between potential parameters and $l$ | The radial wavefunction |
|---|---|---|---|
| 0 | $E_0 = \sqrt{2\omega}(\frac{\sigma}{\sqrt{2\chi}} + 2)$ | $l'(l'+1) + 2\sqrt{4\omega\chi} - \frac{\sigma^2}{2\chi} - \frac{2\sigma}{\sqrt{2\chi}} - \frac{3}{4} = 0$ | $\psi_0(r) = r^{\frac{3}{2} + \frac{\sigma}{\sqrt{2\chi}}} e^{-\left(\sqrt{\frac{\omega}{2}}r^2 + \sqrt{\frac{\chi}{2}}\frac{1}{r^2}\right)} p_0,$ |
| 1 | $E_1 = \sqrt{2\omega}(\frac{\sigma}{\sqrt{2\chi}} + 4)$ | $-2\sqrt{\omega\chi}$ $+\left(2 + \frac{\sigma}{\sqrt{2\chi}}\right)\left(-\frac{l'(l'+1)}{4} - \sqrt{\omega\chi} + \frac{\sigma^2}{8\chi} + \frac{\sigma}{2\sqrt{2\chi}} + \frac{3}{16}\right)$ $+\left(-\frac{l'(l'+1)}{4} - \sqrt{\omega\chi} + \frac{\sigma^2}{8\chi} + \frac{\sigma}{2\sqrt{2\chi}} + \frac{3}{16}\right)^2 = 0$ | $\psi_1(r) = r^{\frac{3}{2} + \frac{\sigma}{\sqrt{2\chi}}} e^{-\left(\sqrt{\frac{\omega}{2}}r^2 + \sqrt{\frac{\chi}{2}}\frac{1}{r^2}\right)}(p_0 + p_1 r^2),$ $p_1 = \frac{1}{4\sqrt{2\chi}}\left(l'(l'+1) + 2\sqrt{4\omega\chi} - \frac{\sigma^2}{2\chi} - \frac{2\sigma}{\sqrt{2\chi}} - \frac{3}{4}\right)p_0$ |
| 2 | $E_2 = \sqrt{2\omega}(\frac{\sigma}{\sqrt{2\chi}} + 6)$ | $\left(-\frac{l'(l'+1)}{4} - \sqrt{\omega\chi} + \frac{\sigma^2}{8\chi} + \frac{\sigma}{2\sqrt{2\chi}} + \frac{3}{16}\right)^3$ $+\left(8 + 3\frac{\sigma}{\sqrt{2\chi}}\right)\left(-\frac{l'(l'+1)}{4} - \sqrt{\omega\chi} + \frac{\sigma^2}{8\chi} + \frac{\sigma}{2\sqrt{2\chi}} + \frac{3}{16}\right)^2$ $+\left(4 - 8\sqrt{\omega\chi} + 2\left(2 + \frac{\sigma}{\sqrt{2\chi}}\right)^2 + \frac{\sigma\sqrt{2}}{\sqrt{\chi}}\right)\left(\frac{\sigma^2}{8\chi} + \frac{\sigma}{2\sqrt{2\chi}} + \frac{3}{16} - \frac{l'(l'+1)}{4} - \sqrt{\omega\chi}\right)$ $-8\sqrt{\omega\chi}\sqrt{3 + \frac{\sigma}{\sqrt{2\chi}}} = 0$ | $\psi_2(r) = r^{\frac{3}{2} + \frac{\sigma}{\sqrt{2\chi}}} e^{-\left(\sqrt{\frac{\omega}{2}}r^2 + \sqrt{\frac{\chi}{2}}\frac{1}{r^2}\right)}(p_0 + p_1 r^2 + p_2 r^4),$ $p_2 = \frac{\left(-8\sqrt{2\omega}\right)p_0 + \left(l'(l'+1) + 2\sqrt{4\omega\chi} - \frac{\sigma^2}{2\chi} - \frac{6\sigma}{\sqrt{2\chi}} - \frac{35}{4}\right)p_1}{8\sqrt{2\chi}}$ |

## Conclusions

In this paper, we have studied the Schrödinger equation for a class of spherically symmetric potentials and illustrated that these models can be treated in a simple and unified manner in the Lie algebraic approach. We have shown that all these models give rise to the same basic differential equation, which is expressible as an element of the universal enveloping algebra of $sl(2)$. We have then obtained the general exact solutions of the basic equation within the framework of representation theory of $sl(2)$ Lie algebra. Also, we have reported the explicit expressions for the energy, wavefunction and the constraint on the potential parameters for each of the systems. The advantage of our algebraic method is that we can quickly obtain the general solutions of the systems for any arbitrary *n*, without the cumbersome procedures of obtaining the solutions for higher states. This method is found to be computationally much simpler than other methods.